\documentclass[12pt]{iopart}

\usepackage{graphicx}
\usepackage{subfig}
\usepackage{color}

\newcommand{\msun}{M${}_{\odot}$}
\newcommand{\hp}{h_+}
\newcommand{\hx}{h_{\times}}
\newcommand{\htl}{\tilde{h}}
\newcommand{\shp}{$h_+$}
\newcommand{\shx}{$h_{\times}$}
\newcommand{\shtl}{$\tilde{h}$}

\newcommand{\Fp}{F_+}
\newcommand{\Fx}{F_{\times}}

\begin{document}
\title[Burst CBC search]{A burst search for gravitational waves from binary black holes}
\author{C Pankow$^1$, S Klimenko$^1$, G Mitselmakher$^1$, I Yakushin$^2$, G Vedovato$^3$, M Drago$^{3,4}$, R A Mercer$^5$ and P Ajith$^{6,7,8}$} 
\address{$^1$ Department of Physics, P.O. Box 118440, Gainesville, FL 32611-8440, USA}
\address{$^2$ LIGO Livingston Observatory, P.O. Box 940 19100, LIGO Lane Livingston, LA 70754, USA}
\address{$^3$ Dipartimento di Fisica, Universit\`a di Padova, Via Marzolo 8, 35131 Padova, Italy}
\address{$^4$ INFN, Sezione di Padova, Via Marzolo 8, 35131 Padova, Italy}
\address{$^5$ University of Wisconsin - Milwaukee, PO Box 413, Milwaukee, WI 53201, USA}
\address {$^6$ Max-Planck-Institut f\"ur Gravitationsphysik
(Albert-Einstein-Institut) and Leibniz Universit\"at Hannover,
Callinstr.~38, 30167~Hannover, Germany}
\address {$^7$ LIGO Laboratory, California Institute of Technology,
Pasadena, CA 91125, USA}
\address {$^8$ Theoretical Astrophysics, California Institute of Technology,
Pasadena, CA 91125, USA}

\eads{\mailto{pankow@phys.ufl.edu}
\mailto{klimenko@phys.ufl.edu}}

\begin{abstract}
Compact binary coalescence (CBC) is one of the most promising sources of gravitational waves. These sources are usually searched for with matched filters which require accurate calculation of the GW waveforms and generation of large template banks. We present a complementary search technique based on algorithms used in un-modeled searches. Initially designed for detection of un-modeled bursts, which can span a very large set of waveform morphologies, the search algorithm presented here is constrained for targeted detection of the smaller subset of CBC signals. The constraint is based on the assumption of elliptical polarisation for signals received at the detector. We expect that the algorithm is sensitive to CBC signals in a wide range of masses, mass ratios, and spin parameters. In preparation for the analysis of data from the fifth LIGO-Virgo science run (S5), we performed preliminary studies of the algorithm on test data. We present the sensitivity of the search to different types of simulated binary black hole waveforms. Also, we discuss how to extend the results of the test run into a search over all of the current LIGO-Virgo data set.
\end{abstract}

\pacs{04.30.-w, 04.30.Tv, 97.60.Lf}
\submitto{\CQG}

\maketitle

\section{Introduction}

The merging of massive compact binary systems is one of the most promising sources for detection of gravitational waves with interferometric detectors. Typically, these sources are composed of binary systems of neutron stars and/or black holes, and the gravitational waves radiated from these sources are coupled to new physics. These sources are critical for a test of general relativity, and provide a valuable way to study the strong-field regime. Their orbits slowly decay because energy is lost in the emission of gravitational radiation, and the frequency of the emitted quasi-sinusoidal gravitational waves increases over time. This semi-periodic stage is called the inspiral. Due to the decay of the orbit, the pair eventually comes close enough for strong-field effects to dominate and no stable orbit is available to the system. The two compact objects then merge together to form a single black hole, releasing a burst of gravitational wave radiation. This stage is referred to as the merger stage. Finally, in the ringdown stage, the newly formed single black hole spins down and radiates gravitational waves in damped quasi-normal modes. The more massive a binary system is, the more energy is expected to be radiated in gravitational waves. Higher mass also corresponds to a lower frequency bandwidth of signal content. These frequencies fall within the most sensitive region for ground based gravitational wave interferometers for stellar mass and some intermediate mass compact binaries.

A number of searches for gravitational waves from compact binary sources have been undertaken in the past, some of which are similar to the search presented here\cite{Acernese:2009sf}. However, most searches\cite{Abbot:2007uq,LSC:2006sf,Abbott:2009bh} are based on the matched filter technique which uses templates for detection of signals of known waveforms\cite{Sathyaprakash:1999cw}. These templates span a parameter space: a set of source parameters such as total mass, mass ratio, spin, and other properties which describe the binary system. It is known that a matched-filter approach is optimal, however, this approach requires the construction of large template banks using accurate theoretical models which is not always possible. Currently, a template bank which covers all possible gravitational wave signals from binary coalescences does not exist. For example, it is expected that some binary mergers will involve components which are spinning\cite{Buonanno:2003nw}. It has been suggested\cite{Shoemaker:2008sf,Vaishnav:2007ws} that coalescing binaries with spinning components will have strong effects on the merger stage of the waveform for masses greater than 50 \msun because of the contribution of higher order modes. While some progress has been made in constructing template banks for the inspiral stage of spinning binaries\cite{Pan:2003br,Buonanno:2004zb}, there are no waveforms describing all stages of coalescence as well as being currently available for use in template based searches.

Despite problems in analytically formulating the late stages of binary coalescences, recent progress in numerical relativity has allowed for a calculation of the waves emitted during the merger stage of two binary black holes. As a result, template banks have been supplemented by numerical simulations of the full inspiral and merger stages of the black hole binary system. This has enabled the construction of high mass ($>$ 35\msun) template banks used in  searches for those binaries. There are still difficulties involved in searching for high mass binaries. The calculation of the numerical waveforms is computationally costly. The generation process makes use of numerical waveforms, but, because of the cost, the waveforms produced are still only approximate in the merger region.

In addition to template based searches, there is another, more general, class of searches that have been carried out for gravitational waves from un-modeled, short duration sources. They are sensitive to a wide class of waveform morphologies and do not require template banks. The detection of un-modeled, or ``burst'', sources is based on excess signal power and correlations of signals detected in different gravitational wave detectors. Because an un-modeled search does not require templates, it is useful in the regions of the compact binary parameter space where there are no waveforms available for use or where the waveforms are difficult to compute. 

In this paper we present a robust method for detection of gravitational waves from compact binary coalescences. This method is expected to be less sensitive than the optimal matched filters, but encompasses a wider class of compact binary sources that may be missed by the template searches.

\section{Search Method}

We first present a brief description of the Coherent WaveBurst algorithm used for searches of LIGO data for gravitational waves from sources which may not be modeled. For a detailed discussion we refer to its implementation details in the references\cite{Klimenko:2007hd}. After the discussion of its un-modeled stage of operation, we will describe a new constraint introduced into the search in the context of detection of binary coalescence signals. We will also outline the execution of a test run using a week of simulated noise data and present the results. These results will include the injection of simulated waveforms into the aforementioned data set, and use those results to estimate the sensitivity of the search to compact binary sources.

\subsection{Coherent WaveBurst}


Coherent WaveBurst is a coherent network algorithm which searches for un-modeled, short duration bursts of gravitational waves in detector strain data using information from all detectors in a network simultaneously\cite{Klimenko:2006io}. Therefore, events are reconstructed for the network, not individual detectors. This is compared to generating detector specific events and attempting to coincide them across detectors \emph{a posteriori}\cite{Arnaud:2003ts}. The algorithm has been used in the fourth\cite{Abbott:2008eh} and fifth\cite{LIGO:2009pz} LIGO science runs in searches for gravitational waves from potential sources such as supernovas, gamma ray bursts, and others. Because of the lack of source models, a statistical approach is employed which is based mainly on excess cross-correlated power inside the detector network\cite{Gusel:1989dd, Anderson:2006dg}. 

In practice, the method described above is performed using time-frequency decomposition. Detector data is read into Coherent WaveBurst and decomposed to multiple time-frequency resolutions. This decomposition transforms a sampled time series into a time-frequency map, with the values of locations on the map representing the energy amplitude in that time-frequency location. Each detector has its own time-frequency map, and in order to create coherent events, the detectors' time series are delayed relative to each other with respect to a set of possible travel times of the gravitational wave. The maps are then combined together for each delay. 

Coherent WaveBurst identifies event candidates as regions on the time-frequency plane which have excess power. A clustering procedure is employed to identify the same time-frequency event across multiple decomposition resolutions. Once all of the clusters are identified, then coherent statistics are calculated for each event candidate.

\subsection{Un-modeled Likelihood}

Coherent WaveBurst conducts the above method in the absence of a source model by use of the likelihood ratio\cite{Klimenko:2007hd}, which has been studied before\cite{Tagoshi:2008ij, Klimenko:2005wa}. A similar framework for detection of the inspiral stage waveforms using the likelihood approach has also been suggested\cite{Pai:2004rf}, but in the context of a template based search. We describe here the construction of a likelihood ratio functional in the model independent case assuming stationary and gaussian noise. We consider a generic waveform $h$ composed of two polarisation states $h_+$ and $h_{\times}$. The effect of a gravitational wave on an interferometer is characterized by a \emph{detector response} $\xi$

\begin{equation}
\xi(\theta,\phi,t,f)=F_+(\theta,\phi)h_+(t,f)+F_{\times}(\theta,\phi)h_{\times}(t,f) .
\end{equation}

The detector response is a function of the sky coordinates $(\theta, \phi)$ through the antenna patterns ($F_+$, $F_{\times}$), which represent the interferometer's sensitivity to the respective polarisation states. In Coherent WaveBurst, the likelihood functional is the ratio of the joint probability $P(x|h)$ of a gravitational wave signal being present in the data to the joint probability $P(x|0)$ of no gravitational wave signal. The ratio is given by:

\begin{equation}
L(\hp, \hx) = \prod_{j}^{N} \prod_k^K \exp\left( \frac{x_{kj}^2}{\sigma_{kj}^2} - \frac{\left(x_{kj}-\xi_{kj}(\hp,\hx)\right)^2}{\sigma_{kj}^2}  \right).
\label{eqn:likelihooddef}
\end{equation}

In the above equation $x_{kj}$ represents the sampled output strain from an individual detector $k$ and time-frequency location $j$. The denominator, $\sigma^2_{kj}$, is the noise variance in the $k$th detector. This is summed over each of the $K$ detectors in the network and time-frequency region composed of $N$ time-frequency samples. This functional is analytically maximized to determine the estimators of $h_+$ and $h_{\times}$ amplitudes for each sample of the data. We consider these amplitudes to be free parameters used in the variation of the likelihood functional. With two (possibly independent) polarisation states, this allows at most $2N$ degrees of freedom to fit with $K\times N$ samples from the data. Because the detector responses are also a function of the sky coordinates via the antenna patterns, the likelihood functional is also maximized over a set of sky locations. The location at the maximum is an estimate of the source location in the sky. In practice, this process involves a large number of free parameters to maximize simultaneously, and the likelihood functional may allow for unphysical solutions for the reconstructed waveforms. In order to mitigate these problems, constraints on the likelihood functional are used to suppress unwanted solutions. In the search presented, we are also using constraints to introduce the model dependence as described in \sref{econs}. The intended purpose of this constraint is to increase sensitivity to compact binary sources as well as better rejection of noise induced events in real detector data.
  
\subsection{Elliptical Constraint} \label{econs}

Since we intend to search for compact binary signals with more definite physical properties than in the un-modeled case, it is possible to further constrain the likelihood functional to better reflect prior knowledge about the signal model. As an inspiralling compact object is a rotating system, the waves produced should be elliptically polarised, and therefore our goal is to enhance the reconstruction of elliptically polarised signals. The enforcement of this constraint should also cause incoherently polarised signals to be poorly reconstructed, which is important in the rejection of glitches. This constraint is expressed by modifying the meaning of the estimators for \shp and \shx:

\begin{eqnarray}
\hp \rightarrow h \nonumber \\
\hx \rightarrow \htl \nonumber \\
\xi \rightarrow \xi' = \Fp(\psi) h+\Fx(\psi)a\htl .
\end{eqnarray}

By this modification, $\xi'$ now represents the detector response to an elliptically polarised gravitational wave composed of $h$ and its ninety degrees phase shifted counterpart \shtl. We will denote all quantities with this phase shift with a tilde. The principal effect of the additional constraint on the algorithm is to reduce number of free parameters required for waveform reconstruction. In the model independent case, there are two independent polarisation states of the gravitational wave. For a set of $N$ sample measurements in the time series, this corresponds to $2N$ free parameters that the algorithm must fit. However, elliptical polarisation demands that one polarisation state $h$ is related to the other \shtl \hspace{1 pt} by a phase shift of ninety degrees. We can exploit this relationship because effectively only one polarisation state is required to be fit, as its phase shifted polarisation state should be identical up to an amplitude factor. The use of this relationship reduces the number of free parameters required to $N$ for $h$ and two additional parameters $\psi$ and $a$. The plane that the states $h$ and \shtl \hspace{1 pt} lie in are rotated with an angle $\psi$ with respect to the detector plane. This angle is the polarisation angle of the wave. Also, the ellipticity parameter $a$ is introduced. This parameter is proportional to the relative strength of the polarisation states where the unity cases ($a=\pm1$) correspond to circular polarisation and the zero case to linear polarisation. For binary sources, this parameter is implicitly related to the inclination angle of the source with respect to the observer.

Coherent WaveBurst makes use of the constraint by substituting $\xi'$ in the definition of the likelihood ratio in \eref{eqn:likelihooddef}. The estimators for $h$ and \shtl\hspace{1 pt} are then obtained in a similar varying procedure as in the un-modeled case.

\subsection{Selection Criteria} \label{cuts}

Usually data collected by gravitational wave detectors is non-stationary and polluted with environmental and instrumental artifacts. Searches use selection criteria to distinguish between gravitational wave signals and ``glitches'' in the data from those artifacts. In previous searches\cite{Abbott:2008eh} using Coherent WaveBurst, there were a set of four statistics, calculated within the likelihood approach, which were chosen for selection of coherent events. We presently describe those statistics, but refer to \cite{Klimenko:2007hd} for a fuller description.

We define the network correlation coefficient $cc$, which tests the overall consistency of the event candidate based on the comparison of reconstructed correlated signal energy $E_c$ and residual noise energy $E_n$. The correlated signal energy is a weighted sum of the correlated energies from all pairs of detectors in the network, and the residual noise energy represents the energy that remains after the reconstructed signal energy is subtracted from the total energy of the event.

\begin{equation}
cc = \frac{E_c}{E_n+E_c} .
\end{equation}

This criteria is intended to exploit the poor reconstruction of glitches. A glitch often can have large residual noise energy as compared to its correlated energy. The network correlation coefficient varies between zero and one, where true gravitational waves should have a value of $cc$ near unity, and glitches will have values of $cc$ significantly less than unity.

We define a penalty factor $P_f$, which is a measure of consistency of reconstructed detector responses. This is a selection criteria for event candidates which have unphysically reconstructed responses. Such candidates have a reconstructed signal energy which is greater than the total energy in a given detector. This statistic is defined by:

\begin{equation}
P_f =  \min_k\sqrt{\frac{\langle x_k^2\rangle}{\langle \xi_k^2 \rangle}}    .
\end{equation}

The quantities $\langle x^2_k\rangle$ and $\langle \xi^2_k\rangle$ are the summed data sample amplitudes and reconstructed detector responses for a selected time-frequency region in an individual detector $k$. If an event is inconsistent, the detector which is maximally inconsistent is chosen. If the event is consistent ($\langle\xi^2_k\rangle < \langle x^2_k\rangle$ for all $k$), then this parameter is defined to be unity.

The next criteria is the energy disbalance $\Lambda$, which is similar to $P_f$ but indicates the overall mismatch between the reconstructed amplitudes and the amplitudes recorded by the detectors:

\begin{equation}
\Lambda = \frac{\sum_k^K|\langle x_k\xi_k\rangle-\langle\xi_k^2\rangle|}{E_c}    .
\end{equation}

It characterizes by the mismatch between the reconstructed energy of the event and the energy of the data. A glitch in one detector can lead to large detector responses being reconstructed in other detectors. For example, the $\langle\xi_k^2\rangle$ term will exceed the $\langle x_k\xi_k\rangle$ term and produce large values of $\Lambda$. Ideally, this statistic should be zero.

All event candidates are assigned a coherent network amplitude. This statistic is the main detection statistic and used to rank events:

\begin{equation}
\eta = \sqrt{\frac{E_ccc}{K}} .
\end{equation}

Because glitches are characterized by a small correlated energy, and therefore can also have a small network correlation coefficient, this statistic is used together with $cc$ to reject various types of glitches. The combination of the two statistics allows the rejection of both consistent but weak glitches, and strong but less consistent glitches. It is used as a detection statistic for event candidates after all other thresholds have been applied. The threshold of $\eta$ used is a function of false alarm rate selected for that analysis.

\Table{\label{tab:eobnrparams} Overview of the parameters used in the injection simulations.}
   	\br
	              & EOBNR\_v1a & EOBNR\_v1b \\	\mr
	total mass (\msun) & $5<M_{total}<100$ & $100<M_{total}<500$ \\ 
	component masses (\msun) & $1<M_i<95$ & $40<M_i<400$ \\  
	minimum mass ratio & 0.012 & 0.1 \\ \br
\endTable

\section{Simulation and Estimation of the Detection Efficiency} \label{sim}

In order to estimate the sensitivity of the search to gravitational waves from compact binary sources, we make use of the EOBNR (Effective One Body Numerical Relativity) family of waveforms\cite{Buonanno:2007sf, Damour:2008lu, Damour:2008sf}. In this model, analytical inspiral-plunge waveforms computed in the Effective One Body formalism\cite{Buonanno:1999sp} are calibrated to the numerical relativity simulations, and matched to the quasi-normal modes of the ringdown to produce complete coalescence signals. The waveforms are only constructed for case of non-spinning black hole binaries.

For estimation of the detection efficiency these waveforms are used by injecting them into the detector strain data, and then the algorithm is applied to detect and reconstruct them. The injections cover a wide range of source parameters. Table \ref{tab:eobnrparams} describes the two sets of injections tested, delineated by total mass. Common between the injection sets were uniform distributions in sky position, inclination angle, and time between injection. All of the waveforms were generated at a distance of 1 
Mpc and scaled to twenty other distances between 100 kpc and 1 Gpc. 

The simulated noise data into which simulated signals are injected has a colored gaussian profile which approximates the LIGO network during its fifth science run. A set of background events are generated from the simulated detector data with no simulated signal injections. These background events are created by shifting detectors in time with respect to each other beyond the maximum possible light travel time of a gravitational wave through Earth. This background set is used to tune thresholds and estimate the false detection rate due to random coincidences by identifying loud background events and fixing the selection criteria thresholds as found in \tref{tab:cuts} to exclude these events from the analysis. The penalty factor, network correlation coefficient, and energy disbalance thresholds are applied to reject loud but inconsistent background events, and the false detection rate ($10^{-6}$ Hz for this analysis) picked then determines the threshold on weak background events via the coherent network amplitude $\eta$.

\Table{\label{tab:cuts} Values of selection criteria used in the analysis.}
		\br 
		network correlation ($cc$) & $>$ & 0.6\0 \\ \mr
		penalty ($P_f$) & $>$ & 0.6\0 \\ \mr
		energy disbalance ($\Lambda$) & $<$ & 0.35 \\ \mr
		coherent network ampltide ($\eta$) & $>$ & 3.0\0 \\ \br
\endTable

Once this stage is complete, then the simulated compact binary waveforms are injected into the data, and the algorithm is applied to detect these. Simulated event candidates are passed through the same set of selection criteria which were applied to the background.

\subsection{Simulation Detection Results}

\begin{figure}[htbp]
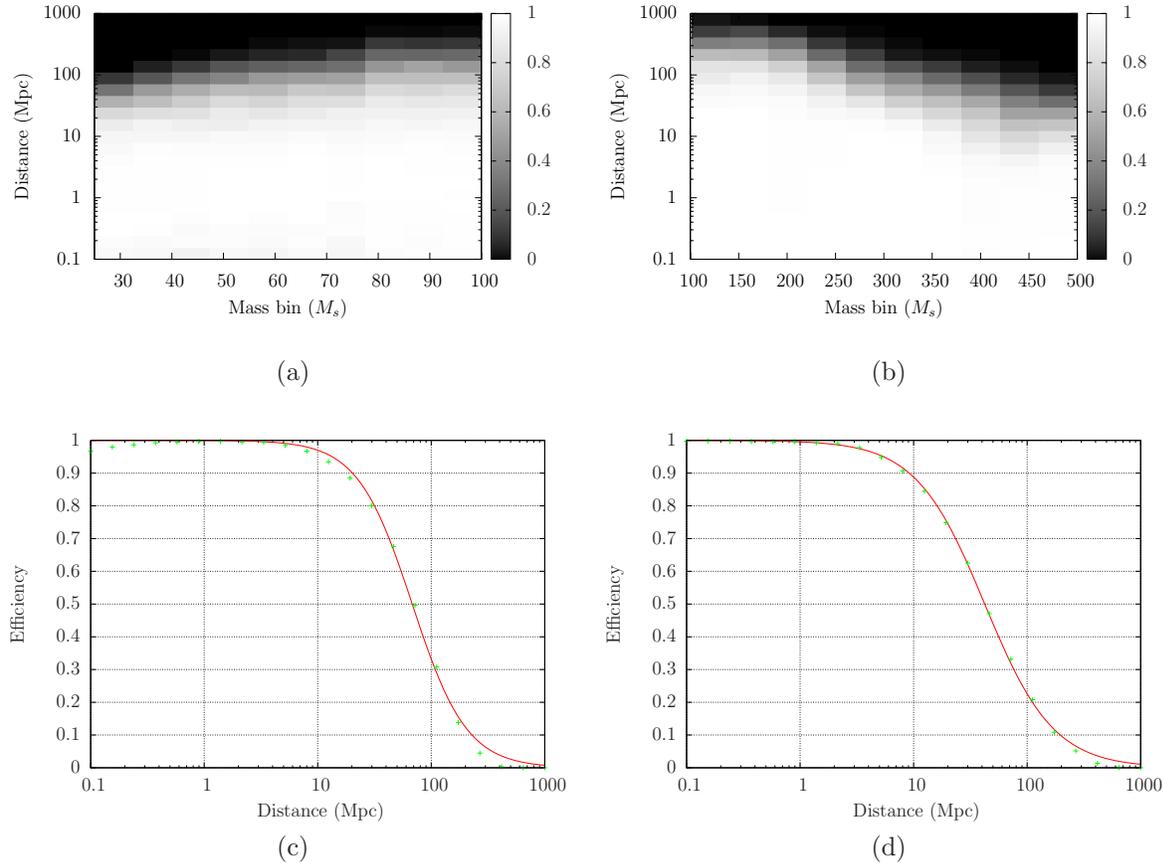
%
   \subfloat[]{%
      \scalebox{0.6}{ \input{efficiency_low_figure_1a.tex} }
      \label{fig:effl}
   }
   \subfloat[]{%
      \scalebox{0.6}{ \input{efficiency_high_figure_1b.tex} }
      \label{fig:effh} \\
   }
   \\
   \subfloat[]{%
      \scalebox{0.6}{ \input{total_eff_low_figure_1c.tex} }
      \label{fig:teffl}
   }
   \subfloat[]{%
      \scalebox{0.6}{ \input{total_eff_high_figure_1d.tex} }
      \label{fig:teffh} \\
   }
   \caption{Network detection efficiencies for the low mass (left) and high mass (right) sets. The top plot shows efficiency over the total mass and injection distance of the system while the bottom plot is the efficiency after integration over the total mass dimension.}
   \label{fig:eff}
\end{figure}

Estimated detection efficiencies are shown for both sets of waveforms in \fref{fig:eff}. From \fref{fig:teffl} and \fref{fig:teffh}, it is shown that we are sensitive to the wide range of parameters exhibited by these injection sets. In particular, our detection efficiencies at close distances approach unity, and therefore we are sensitive to the entire set of mass ratios used in the injection sets. It is also evident from \fref{fig:effl} and \fref{fig:effh} that we are sensitive to a wide range of total system mass, enabling investigation of both stellar mass black hole binaries and some intermediate mass binary black hole systems. Also, almost all of the total mass bins in figures \ref{fig:effl} and \ref{fig:effh} have a 90\% detection efficiency up to 20 Mpc. This distance encloses the Virgo cluster, an abundant collection of potential hosts for sources. 

\subsection{Estimation of Visible Luminosity}

We estimate the number of hosts visible to the search, via their luminosity, by use of the network detection efficiencies and a catalog of galaxies. The catalog compiled by the compact binary coalescence group in LIGO\cite{Kopparapu:2007kx} with luminosity entries is used because of the unknown relationship between compact binary source populations and their host galaxies. A simple relationship between the stellar birth rate, as measured by the blue light luminosity of a galaxy, and the binary coalescence rate is assumed. Because of this simplification, luminosities calculated by compact binary searches are typically calculated in units of blue light luminosity $L_{10}$. This luminosity represents the output of $10^{10}$ G type stars in the blue light (445 nm) optical filter\cite{Binney:1998qm}. For reference, the Milky Way galaxy is approximately $1.7\,L_{10}$. The total luminosity visible to a search, as measured in units $L_{10}$, is called the cumulative luminosity. The cumulative luminosity is a measure of the effective number of galaxies surveyed by a search and is used in rate calculations in template based searches. We calculate the efficiency $\epsilon$ of the network and use this to convolve with each galaxy having a location $(r_i, \theta_i, \phi_i)$ and luminosity measurement $L_i$:

\begin{equation}
C_L = \sum_i \epsilon(r_i, \theta_i, \phi_i) L_i .
\end{equation}

In this study, we have considered a wider range of masses than previously examined in other searches, and as a consequence the increased gravitational wave energy output for higher mass binaries and the corresponding increase in search volume available allows for an order of magnitude increase in detected luminosity. In practice, the cost of a widening the mass range is reflected in the subsequent requirement of less restrictive selection criteria, which leads to a higher false detection rate. However, the simulated nature of the data does not contain those transients, which tend to form the outliers in the distribution of $\eta$, and we are forced to disregard the effects on the false detection rate. As a consequence of using simulated data, the threshold on $\eta$ is decreased and will therefore lead to comparatively better efficiencies. Matched filter searches using the most current set of LIGO data obtained a cumulative luminosity of $5500\,\textrm{L}_{10}$ for binary black hole systems of total mass less than 35 \msun\cite{Abbott:2009bh}. In our test with simulated data, we estimate a cumulative luminosity of above $22000\,\textrm{L}_{10}$ for the low mass injection set. The matched filter cumulative luminosities quoted are for searches involving the Livingston detector and both Hanford detectors. 

\section{Future Work} \label{futurework}

There are a number of items that have not been addressed in the test run presented, but will be included in the full search. Because the test run used simulated noise, rather than actual detector data, we expect that some of these estimates are optimistic. There are a number of issues involved in actual detector data. One of the more difficult issues to deal with are instrumental and environmental transients falsely identified as event candidates, however, the enhancement of our selection procedures with regards to the elliptical constraint should be able to mitigate this problem. More study should allow us to characterize how well we can reject glitches in real detector data, and therefore how effective the elliptical constraint is in the reduction of glitches.

The latest data taking run by LIGO has produced the equivalent of a year of data in which all three interferometers were taking science quality data simultaneously. Additionally, in the later part of this run, the Virgo interferometer was also in coincidence with the LIGO network and the data taken from the Virgo run will be analyzed jointly with LIGO data. It is also possible to perform more extensive simulations with these and other waveforms\cite{Ajith:2007fj} in order to place plausible upper limits on the rate of compact binary coalescences in the universe in the event that gravitational wave signals are not confidently detected.

It is expected that LIGO and Virgo will start a joint run with upgraded detectors before the end of 2009. The search presented can also be used for the analysis of this new, more sensitive data.

\section{Summary}

We have presented some of the motivations for carrying out a search for compact binary coalescences with an un-modeled technique. These motivations include detecting events which may not be well modeled analytically and therefore may not be detected by a template based search. To this end, we have outlined the plans for a search for binary coalescences using a weakly-modeled search algorithm which has been modified to better detect the intended source model of the search. In this regard, the search is intended to be a complement to existing template based searches. These modifications are intended to improve the reconstruction of elliptically polarised gravitational wave signals. We have also presented the results of a test run using simulated LIGO data and simulated binary coalescence waveforms in order to estimate the efficiency of the search in detection of binary coalescence signals. We have demonstrated the ability to detect 90\% of injected signals at 10 Mpc for systems of total mass between 100-500 \msun and 20 Mpc for systems of 25-100 \msun. 

\ack

We gratefully acknowledge helpful discussions with Evan Ochsner, Ray Frey, and Chad Hanna in the development of this paper. This work is funded by grants PHY-0555453 and PHY-0653057 from the National Science Foundation. This paper has been assigned the LIGO document number P0900062.

\section*{References}

\bibliographystyle{unsrt}
\bibliography{references}

\end{document}